# Building Metaverse Responsibly: Findings from Interviews with Experts

*Research Full-length paper*

**Muhammad Irfan Khalid**
University of Agder, Norway &
Singular Logic, Greece
muhammad.i.khalid@uia.no
mikhalid@singularlogic.eu

**Ilias O. Pappas**
University of Agder & Norwegian
University of Science and
Technology, Norway
ilias.pappas@uia.no

**Moatasim Mahmoud**
Singular Logic, Greece
mmoatasim@singularlogic.eu

**Stamatia Rizou**
Singular Logic, Greece
srizou@singularlogic.eu

## Abstract

The metaverse promises unprecedented immersive digital experiences but also raises critical privacy concerns as vast amounts of personal and behavioural data are collected. As immersive technologies blur the boundaries between physical and virtual realms, established privacy standards are being challenged. However, little is known about how the experts of these technologies (such as requirement analysts, designers, developers, and architects) perceive and address privacy issues in the creation of metaverse platforms. This research aims to fill that gap by investigating privacy considerations in metaverse development from the expert's perspective. We conducted in-depth, semi-structured interviews with metaverse platform and application experts to explore their views on privacy challenges and practices. The findings offer new empirical insights by extending information systems privacy research into the metaverse context, highlighting the interplay between technological design, user behaviour, and regulatory structures. Practically, this work provides guidance for developers, and policymakers on implementing privacy-by design principles, educating and empowering users, and proactively addressing novel privacy threats in metaverses.

## Introduction

The term "*metaverse*" combines "meta" (beyond) and "verse" (universe) (Dolata & Schwabe, 2023), and has garnered significant attention in recent years, with searches increasing by 7,200 percent and mentions in U.S. regulatory filings rising dramatically in 2021 (McKinsey, 2022a; Ball, 2022). The metaverse refers to an interconnected, immersive, and persistent virtual environment where users interact with each other, digital objects, and the surrounding virtual world via avatars, facilitated by technologies like virtual reality (VR), augmented reality (AR), and spatial computing. This virtual space contains a wide array of digital spaces, platforms, and experiences, enabling real-time social, economic, and cultural interactions while mixing physical and virtual realities. As a result, it raises new and complex privacy concerns due to the massive collection of personal, behavioural, and interaction data within these digital spaces.

Due to the intrinsic immersive features of the metaverse technologies, the user behaviour is evident, with sensitive data such as facial features, eye movements, and even physiological reactions being collected (Dwivedi et al, 2022), which can unintentionally expose intimate personal details such as those of health status and personality traits (Bao et al, 2022). In the same manner, such highly sensitive data are not only accessible by service providers but also by other users, elevating the risks of harassment, stalking, and other privacy intrusions (Falchuk et al, 2018). Although such issues and related risks have been discussed, the existing privacy standards and regulations are often inadequate to address the novel and complex privacy challenges posed by immersive digital environments (McKinsey, 2022b).

The existing research on privacy in digital platforms has primarily focused on traditional online environments or virtual worlds (Boughzala et al, 2012). However, as the metaverse evolves, these studies are no longer sufficient to address the new privacy issues encountered by immersive and persistent virtual spaces. Although studies have begun to explore privacy issues in metaverse like environments (Dincelli & Yayla, 2022), they often overlook the expert perspectives on how these challenges are identified and mitigated in the design and development of metaverse platforms. In particular, there is a gap in understanding how professionals involved in metaverse creation such as requirement analysts, designers, developers, and architects perceive and address privacy issues in their work.

This research aims to fill this gap by investigating privacy considerations in metaverse development from the perspective of industry experts. Specifically, this study explores the privacy challenges faced during the creation of metaverse platforms and the strategies employed by developers and designers to address these concerns. The research contributes to new insights into the field of information systems by extending privacy studies into the metaverse context and by providing empirical data on how privacy risks are managed within the development lifecycle. The findings aim to offer both theoretical contributions to privacy research and practical guidance for developers and policymakers to implement privacy-by-design principles, educate users, and address emerging privacy threats. Our study is guided by the following research question:

**RQ:** *How do experts in the design and development of metaverse platforms perceive and address privacy risks during the creation of these immersive environments?*

We have addressed this research question through in-depth, semi-structured interviews with metaverse platform and application experts. By analyzing their responses, we gained insights into the practices, challenges, and potential solutions for managing privacy in metaverse development. The findings of this study contribute to both academic literature and practical strategies for industry professionals and policymakers, ensuring privacy concerns are effectively addressed in the design and regulation of metaverse platforms.

The rest of the paper is organized in the following way: Next section focuses on the background of related research studies in the metaverse privacy context. Next, we describe the methodology. Subsequently, we present the results. Finally, we provide discussion with synthesis, implications for future research and with limitations and future research directions.

## Background

The metaverse, an interconnected digital space that merges with AR, VR, and other immersive technologies, has garnered significant attention in recent years for its potential to revolutionize a wide range of industries, from entertainment and marketing to healthcare and education (Peukert et al, 2022). As these technologies advance, they promise to create more engaging and interactive digital environments, transforming how users interact with data and one another. Despite the growing enthusiasm for these innovations, current research has mainly focused on theoretical and conceptual discussions surrounding their applications, with far less importance on mitigating the ethical issue that are posed by these immersive technologies in real-world settings (Khalid, 2025). Particularly, the ethical implications, mainly those related to end user's data privacy and protection, have not been sufficiently studied, notably from an empirical perspective.

The existing literature has highlighted the theoretical foundations and prospect of metaverse technologies by researching topics like user experience enhancement and commercial opportunities (Wang et al, 2022). However, these contributions usually overlook the ethical, practical, and regulatory challenges that accompany the development and deployment of such immersive systems. In this existing literature on privacy in metaverse technologies, systematic reviews and conceptual frameworks (Tiwari et al, 2025;

Marabelli & Newell, 2023; Nair et al, 2022) have shed light on the potential of the metaverse, but they rarely delve into how developers can proactively address and perceive privacy and security concerns in the design and implementation stages.

The rapid evolution of AR and VR technologies has also presented significant problems regarding the collection, storage, and processing of user data. Metaverse systems typically gather vast amounts of data, including user's movements, preferences, interactions, and even biometric and behavioural data, raising crucial privacy issues (Dincelli & Yayla, 2022). While several studies have highlighted the risks of data misuse and privacy breaches in these environments, most focus has been on theoretical privacy concerns rather than the real-world challenges faced by developers who are responsible for assuring that these technologies comply with privacy standards and regulations (Nair et al, 2022). For example, one such study by Nair et al, (2023) and others has shown that seemingly harmless data points, such as user's movements or interactions in virtual spaces, can be used to infer sensitive personal details, such as psychological profiles or physical health information. This emphasizes the need for developers to address privacy from the outset of the design process, rather than as an afterthought.

The gap in the literature is particularly evident when examining the role of developers, designers, and other industry professionals in integrating privacy protections into the metaverse. While some scholars have examined privacy risks from a high-level perspective (Peukert et al, 2022; Wang et al, 2022), few have provided practical guidance on how developers can implement safeguards within the infrastructure of metaverse systems. This lack of focus is significant since privacy concerns are not merely theoretical but must be actively managed throughout the entire development lifecycle. For instance, Nair et al, (2023) highlight the risks of unique user identification based on motion data but fail to address the practical challenges developers face in anonymizing or securing such data.

Also, existing frameworks like GDPR give some regulatory guidance, while their application to metaverse platforms remains underdeveloped. The challenge lies in decoding these high-level privacy frameworks into actionable measures for developers by guaranteeing that data protection is entrenched in the very design of metaverse technologies (Smith et al, 2011). As pointed out by Floridi (2020) and Zuboff (2019), privacy and data sovereignty are critical concerns, but the focus has largely been on theoretical discussions rather than the practicalities of building secure systems in the metaverse.

The lack of clear data governance frameworks for metaverse environments compounds the emerging ethical challenges. While there has been significant philosophical discourse on topics like surveillance capitalism (Zuboff, 2019) and privacy as contextual integrity (Nissenbaum, 2004), there remains a notable gap in empirical research on how these issues manifest in the development and use of metaverse technologies. Al-Kfairy et al (2025) and Chen et al (2024) have begun to address these concerns by discussing the ethical and privacy challenges in the metaverse. Still, they often lack detailed, actionable insights into how professionals on the ground can mitigate these issues. Their frameworks, though valuable in identifying ethical dimensions, do not provide sufficient practical guidance on how to integrate privacy safeguards during the technology development process.

Given the substantial risks associated with user data in the metaverse, there is a critical need for empirical research that studies the real-world ethical dilemmas faced by developers, regulators, businesses, and users. As shown by Nair et al (2023), even short periods of data collection can result in exact identification of individuals by raising concerns about consent and data ownership. Such studies highlight the importance of including privacy protections from the early stages of a metaverse's system design.

In order to handle these concerns, this paper offers an empirical investigation into the ethical implications of metaverse technologies, particularly focusing on privacy risks and the responsibilities of developers, businesses, and regulators in ensuring ethical practices. Drawing from in-depth semi-structured interviews with industry experts, including designers, developers, and business analysts, this study seeks to gain insights into the challenges and opportunities in designing privacy-conscious, responsible metaverse platforms. By examining the perspectives of those directly involved in the development of immersive technologies, this research aims to fill the gap in current literature by offering both theoretical and practical recommendations on how ethical concerns, particularly privacy issues, can be effectively addressed in the evolving metaverse landscape. The results of this study are essential for understanding how a responsible metaverse can be developed, one that not only promotes innovation and user engagement while also maintains ethical standards in data collection, privacy protection, and user autonomy. Since immersive technologies continue to advance, it is vital to handle the ethical challenges that come with these innovations, confirming that user privacy is protected and that the technology is designed in a way that satisfies all stakeholders.

## Research Method

### *Stage 1: Collection of data by in-depth semi-structured interviews:*

Metaverse experts are defined as those industry professionals that are involved at any stage of metaverse technologies/applications creation such as, requirement engineer, designer, developer/programmer/, product architects, team leaders, business development managers, and managers. They can also be termed as "creators of immersive technologies/platforms". To find these experts, we searched over LinkedIn, searching open ended over the internet, using some company's connections. We managed to identify 40 professionals working in one of the above-described roles. The identified individuals were contacted and asked to participate in this research. Twelve individuals agreed to participate in this research working on various roles for metaverse technologies/applications creations. Details of those metaverse experts are provided in Table 1. Interview participants have diverse expertise from metaverse requirements engineer, to programmer, backend engineer, XR decision lead, business development manager, and VR developers. The interviews were conducted in English face to face using online video conferencing software. We have performed semi-structured interviews to gauge all the required details from the interviewees. These interviews were conducted from December 2024 until June 2025 and last between from 45 minutes and 1 hour.

| ID | Position | Metaverse Experience | Company Type | Years of Metaverse Experience | Company Size | Company Location |
|---|---|---|---|---|---|---|
| P1 | University Professor | XR | Higher Education, Human-Computer Interaction and Extended Reality (XR) Development | 8 | Academic Research Group | Italy |
| P2 | Software Architect | VR | Metaverse & Extended Reality (XR) Solutions | 9 | 2–10 employees | Greece |
| P3 | Business Development Manager | VR | Metaverse & Extended Reality (XR) Solutions | 2 | 2–10 employees | Greece |
| P4 | Product Owner | XR Specialist | Maritime Research and Development (R&D) | 10 | 201–500 employees | Netherlands |
| P5 | Software Engineer | VR applications | Technology/Software Development, VR/AR, Entertainment | 11 | Small-to-Medium | Cyprus |

| P6 | Software Developer | VR applications | Research and Development (R&D) in Interactive Media and Emerging Technologies | 4 | Medium to Large | Cyprus |
|----|----|----|----|----|----|----|
| P7 | Software Developer | VR | Research and Development (R&D) in Digital Cultural Heritage | 10 | Medium to Large | Cyprus |
| P8 | System Engineer | XR | Research and Development (R&D) in Interactive Media and Emerging Technologies | 2 | Medium to Large | Cyprus |
| P9 | Requirements Engineer | AR, VR | IT / Digital Transformation | 2 | 1,001–5,000 employees | Greece |
| P10 | Manager & Decision Lead | XR | R&D in Electrical and Computer Engineering | 4 | Over 500 employees | Greece |
| P11 | Software Designer | XR | IT / Digital Transformation | 6 | 1,001–5,000 employees | Greece |
| P12 | Software Engineer | XR | IT / Digital Transformation | 7 | 1,001–5,000 employees | Greece |

**Table 1. Metaverse Expert's Details**

The purpose of these interview was to understand the perceptions of metaverse actual creators regarding ethical considerations like their own view of privacy in the metaverses, privacy related to users data, critical privacy issues in the metaverses, their current practices that have been in use to ensure ethics considerations such as privacy by design, challenges that have been faced when it comes to ensure data privacy, transparency, how much current regulations such as GDPR being followed while ensuring subject's data privacy, and informed consent policies. The interview guide was developed after reviewing the literature practically related to "ethical considerations in metaverse technologies". The interviews were open-ended in nature and there were no right, or wrong answers expected. Based on the insights provided by the interviewee, we asked probing questions to have more detailed insights, probing questions was a follow up question. For instance, in several interviews, we offered interviewees to provide further examples in order for them to clearly understand the question and provide detailed insights into those questions.

In total the interview guide consisted of 15 questions, the first part was about asking the demographic details, their background in terms of working with metaverse technologies, their current position, experience related to metaverse technologies, and their experience in working with the creation process of metaverse technologies. Part 2 was about their own definition of privacy in metaverse technologies, difference of privacy in conventional 2D application and those of in metaverse 3D applications, privacy risks that are most critical in metaverses. Part 3 was the most important part of the interview that was about how their organization approach privacy compliance while creating metaverses technologies, framework which is being used to ensure privacy protection, challenges while adhering to certain regulations such as GDPR, how they approach the trade-off between immersion and protection, do they involve users while creating these metaverse applications, their understanding about privacy by design approach, and which metaverse technology is most dangerous for privacy violation. Part 4 was about the impact of privacy on user trust, engagement and data sharing, their perceptions about the technical organizational, and legal requirements about ensuring privacy in these metaverse technologies, and user's behaviour towards their privacy in metaverses.

### Stage 2: Data analysis:

All interviews were transcribed verbatim and analyzed using MaxQDA. The research team followed the Gioia methodology (Gioia et al, 2013), which involves a three-tier coding process: first-order concepts, second-order themes, and aggregate dimensions. In the first phase, we used open coding to extract first order concepts directly from participants own words. These concepts were then clustered into higher level second order themes, capturing broader patterns and relationships within the data. In the final phase, we synthesized these second order themes into aggregate dimensions, which represented the overarching theoretical categories that emerged from the analysis. This rigorous, systematic approach ensured a detailed and meaningful interpretation of the data, grounded in the participants experiences while also allowing for a conceptual understanding of the ethical considerations under study.

## Results

### User awareness, empowerment, and trust

The first dimension is based on the user's perspective about the metaverse privacy. Developers are concerned about user's low awareness of privacy, the directive to empower users with control, and the connection between privacy and user trust. This dimension encloses three second-order themes: (1) low user privacy awareness and informed consent, (2) empowering users through privacy controls dashboards and transparency, (3) privacy as an element in user trust and engagement.

Developers overwhelmingly reported that many users are not fully aware of privacy issues in the metaverse and often give consent without understanding. Interviewees noted a *"general lack of user awareness"* about what data is collected, how it's used, and the implications of immersive virtual environments. For example, one participant pointed out that *"you have the terms and conditions for metaverse platforms that nobody reads, you basically agree to that, and you don't know, or you don't care,"* (P10). Lengthy, legalistic privacy policies discourage reading, leading to uninformed consent. As another developer put it, *"if I read the terms and conditions of Meta's metaverse offerings, I cannot read them. To be honest, nobody can,"* (P9). Users often underestimate the risks or see privacy loss as *"an inevitable trade-off"* for using the metaverse, a space where personal data is more intimately tied to virtual experiences and social interactions. Several participants described this attitude as a form of pragmatic resignation, where people assume nothing bad will happen despite the greater vulnerability in a virtual world. This theme highlights an ethical concern that user's ignorance in the metaverse leaves them vulnerable, placing greater responsibility on developers to protect privacy by default in immersive, digitally constructed environments.

Many interviewees stressed that user-facing privacy features like clear consent mechanisms, granular settings, and transparency about data use are critical for building trust in metaverse platforms. *"The more controls you give me in the metaverse, the bigger the chance I trust the platform,"* (P12) explained, suggesting that *"when users can opt in/out and configure what personal information is shared in these virtual spaces, they feel safer,"* (P7/P10). Participants gave examples of implementing toggle switches, permission prompts, and dashboards that let individuals decide what aspects of their digital persona and virtual activities they want to share. In one case, when a team added a prompt asking users *"Do you agree to share your information in this metaverse environment, virtually everyone declined, none of them opted in,"* which unambiguously demonstrated that given a real choice, users will withhold data (P9/P12). Developers mentioned offering anonymous or incognito modes as another empowering feature in the metaverse. As one interviewee reminisced, *"You should be able to join a VR game in the metaverse and say, OK, I don't want to log in, have nothing being monitored because I chose not to,"* recalling the *"good old times"* when virtual experiences didn't come with constant tracking (P10). Such design choices (e.g., allowing guest access with no persistent profile) minimize data collection by default, ensuring privacy even in the most immersive spaces.

Many interviewees asserted that strong privacy protection is directly linked to user trust in a metaverse platform. "*If you learn down the line that a company in the metaverse collected your data and then sold it, that happens a lot such that it leaked user information, leaked virtual world activity, you [lose] trust in the application,*" (P12). Several developers expressed that a single privacy violation or scandal in the metaverse could permanently damage an application's reputation. Conversely, visible privacy features reassure users and encourage them to continue using the service. In the words of one interviewee, "*It does affect users and whether you trust the application especially if you find out later what they were doing with your data in the metaverse.*" Here, trust is seen as cumulative, built over time by honest practices in virtual worlds but "*easily broken by a single violation.*"

Importantly, some participants offered a nuanced view that trust depends not just on user interface controls but also on a company's broader reputation and behaviour in the metaverse. One expert said he "*would not care if you give me opt-in/opt-out for every kind of data stream in the metaverse, I would care more about who the provider is and what trust level they have earned*", based on their track record (e.g., news reports, legal cases) (P8). In other words, a platform well-known for respecting privacy in its virtual environments will be given a margin by users, even if it collects some data, whereas a company infamous for breaches in the metaverse will not be trusted despite offering checkboxes. Users eventually determine if a privacy-friendly interface is genuine or just a deception. As a result, designers must consider privacy at two levels: the interface level (providing user options and consent in immersive virtual spaces) and the institutional level (ensuring the company truly handles data responsibly and communicates its values in a metaverse context).

### Privacy by design and developer (s) responsibility

The second dimension revolves around the creation/development of the metaverse technologies. Specifically, there is an obligation to proactively integrate privacy into the design phase at the time of products early development, and there are considerable practical challenges developers face in doing so. This dimension combines underlying themes about internal practices such as (1) privacy-by-design principles (integrating privacy early), (2) challenges in implementing privacy protections, and (3) strategies and best practices for ethical data handling. Together, these reflect the ethical stance that developers must build privacy in "from scratch" and the real-world difficulties and solutions in achieving that goal.

A strong recurring theme was that privacy should be ingrained from the very start of metaverse software projects, rather than added retroactively. Many participants argued that privacy requirements must be considered during initial planning and architecture by treating privacy as a fundamental design criterion. "*These features should be coming right from the start,*" one expert insisted, meaning privacy can't be an afterthought once core features are done (P5). In practice, this means creating data flow diagrams early in development and identifying points where personal data is collected, stored, or transmitted. One participant advised that if a piece of data is not truly necessary, it's better not to collect it: "*If there's a way to skip [collecting a piece of data] entirely ensure there's no privacy concern there*". This proactive data minimization mindset is a fundamental principle of privacy by design techniques.

Several interviewees emphasized that the first-time real users interact with the system, privacy protections should already be in place. One developer from a research context explained that as soon as they move from internal prototypes to any external user testing, they must do everything "*by the book*" regarding data handling. "*The moment you engage users you need to have privacy in it at the very first iteration,*" (P11) said, stressing that even in fast, agile development cycles, one cannot wait until later versions to implement consent forms, anonymization, or security measures. Thus, privacy by design was described as "*something important that will start from scratch,*" and it should be ingrained in the project's DNA just like any core feature or performance requirement (P9). Interviewees also touched on the organizational aspect of privacy by design: "*Before you start, put some main [privacy] milestones. Based on that feedback you decide where you need checks for privacy or if there's a way to ensure there's no privacy problem at all,*" (P2).

Despite best intentions for privacy by design, developers confront numerous challenges when turning these principles into practice. The interviewees candidly described technical, organizational, and even psychological hurdles that can impede privacy implementation in metaverse projects. One major challenge is the complexity and opacity of data flows in modern applications. Metaverse systems often consist of many components making it hard for developers to even know all the data being collected and where it goes. *"It's not easy every time to find where data is collected and by what component,"* (P12) explained, highlighting that complex software architectures obscure data pathways. Occasionally, third-party libraries are integrated that are not familiar to the team are quietly collecting user data from the background. *"Sometimes you use third-party libraries, and these libraries collect some data you're not aware of. So, finding all these parts is quite tedious,"* the same interviewee noted.

One participant pointed out that a rushed developer might *"not hash the passwords or not use authorization tokens"* if corners are cut, leaving data exposed (P3). A lead engineer admitted, *"I have to check on my own that those practices are followed,"* indicating that without rigorous code reviews and oversight, privacy best practices might be skipped in the rush (P7). Compliance with external platforms and evolving standards is another headache. One developer gave the example of Oculus (Meta's VR platform) issuing an update that introduced new privacy rules, forcing developers to react: *"You see a lot of updates in the Oculus platform, where you need to go back and see what has changed… if they want to store some specific location info, you need to… re-structure your program… comply with those rules no matter what."* (P2). Compounding this, interviewees lamented the lack of official guidelines or industry standards tailored to XR privacy. *"Currently we don't have any state-official guidelines or best practices for VR privacy,"* (P2).

### Regulatory and governance environment

The third dimension captures the developer's concern regarding the external regulatory and policy context that impacts metaverse privacy. This includes the adequacy (or inadequacy) of current data protection laws, the need for new guidelines specific to immersive technologies, and how government or industry governance can shape privacy outcomes. In essence, this dimension is about how top-down forces (laws, regulations, standards) are perceived by developers in addressing ethical issues. It comprises themes of (1) gaps in current data protection laws for metaverse, (2) the need for evolving regulations and standards, and (3) the role of regulation in user awareness and industry practice.

A prevailing sentiment among developers is that existing privacy laws (like the EU's GDPR or similar frameworks) are necessary but not sufficient for the metaverse. Interviewees noted that while laws like GDPR provide a strong baseline for personal data protection, they did not anticipate many of the unique data types and scenarios present in AR/VR. *"GDPR doesn't say anything about these 3D applications. It says any type of application that collects data that can lead to identification of a person. But if a metaverse application is collecting your surroundings, your house. I don't know if it can identify the person, but it tells a lot about the person's habits,"* one participant explained (P12). This highlights a gap: spatial mapping data or biometric gaze data might not directly identify an individual by name, yet they reveal highly personal information (one's living environment, daily routines, emotional states, etc.). Current laws generally categorize personal data broadly, but they don't explicitly address things like continuous sensor streams, 3D environmental scans, or physiological signals that are central to XR. Developers feel that these nuances of "dynamic" and immersive data aren't fully captured by traditional legal definitions of personal data.

As a result, many interviewees argued that new or updated regulations are needed to cover XR-specific privacy issues. Some called for *"rethinking regulations"* to provide *"more detailed, more nuanced, more front facing"* rules tailored to immersive technologies (P11). There is an expectation that legislators will eventually catch up to XR, as they usually do with tech advancements (albeit with a lag). One optimistic view was: *"Legislators are always late to catch up with technology, but eventually they will. Down the line it will be fused into one piece of legislation for all kinds of technology applications,"* predicting an overarching law that includes XR (P12). Others thought maybe a metaverse-specific framework or at least official guidance on VR/AR data would be necessary. Concrete suggestions included something similar to today's web cookie consent banners but for VR: e.g., a standardized privacy onboarding form in any AR/VR app that clearly discloses what will be collected (an interviewee likened it to *"applying something like the GDPR form before using anything in the Metaverse"*) (P2). Such ideas reflect a desire for uniform practices that all XR applications would adhere to, making privacy expectations clear and consistent for users.

In the meantime, until new laws materialize, developers often lean on existing regulations (like GDPR) and general best practices as guidance. Many mentioned that if their application has users in Europe, they *by* *default ensure GDPR compliance* for those users. One developer recounted the effort required: *"When we had to publish the application and be GDPR compliant, it took a lot of effort, we had to change features or develop new ones,"* (P12). For instance, they implemented an *"account deletion option and avoided collecting certain data"* specifically to meet GDPR requirements. This shows that regulations can directly shape product features legal mandates like data portability, right to delete, or explicit consent get translated by developers into concrete design choices. Many interviewees indicated they work closely with legal teams or privacy lawyers to interpret broad laws for their specific context. *"We are in contact with our lawyer they tell us what we should do to be compliant,"* (P8) said one startup developer, highlighting how legal counsel informs technical decisions like what data to ask for, how to phrase privacy policies, and how to store data in accordance with health or finance data rules if applicable. This reliance on legal interpretation underscores that while laws set principles, implementing them in a cutting-edge XR app requires careful translation.

A complexity developers pointed out is that metaverse platforms are global, but laws are regional. GDPR was frequently cited as a gold standard, but as one noted, *"GDPR applies only to Europe there's nothing stopping me from accessing the metaverse in another country,"* (P11). This raises jurisdictional questions: should a metaverse service apply GDPR-level protections to all users worldwide, or only EU users? What about countries with weaker privacy laws, do companies lower the standard there, or maintain one high bar? Some respondents hinted that forward thinking companies choose to implement the strictest standard globally for simplicity and ethical consistency, although it was not universally stated. In this regulatory environment, most developers prefer to rely on caution, relying on known frameworks and keeping away from any new developments. One person mentioned they *"run everything through a legal paper from day one"* to cover their bases, implying they formally document compliance and legal review as part of the project kick-off (P2).

### Novel ethical risks of metaverse data

The fourth aggregate dimension encompasses the unique privacy risks and ethical challenges that arise from the nature of metaverse technologies. In developers understanding, metaverse platforms introduce new vectors of data collection and potential abuse far beyond traditional apps. This dimension apprehends themes related to (1) unprecedented scope of personal data in the metaverse, (2) new forms of privacy invasion and harm, and (3) balancing innovation with privacy protection in the future. It outlines the potential concerns that developers have not only about what is already problematic, but also about what could become major ethical issues as the metaverse evolves.

A fundamental difference in the metaverse is the breadth and depth of personal information that can be collected. Unlike a 2D website that might log clicks or text input, an AR/VR experience can potentially record a person's physical environment, biometric signals, and behavioural patterns in fine detail. One participant vividly noted, *"the metaverse has the complete map of your home... you walk around in your home playing... they have a 3D spatial map of your entire house. From that they can do calculations – come up with assumptions about your income, for example, and other stuff." (P12).* This example shows how seemingly innocuous data can enable inferences about your lifestyle or socioeconomic status (perhaps based on furniture, layout, etc.).

*"For a lot of things, we consider immersive, I need to have a lot of sensors gathering information about a human. That's the part I consider critical,"* (P11). Body language and motion data are another new category that *"are not generally available in mobile phones or desktops... this is something very new,"* (P7/P8). These behavioural patterns can act like unique identifiers *"how a person walks is unique, how they speak is unique, their body language is unique"* meaning that even without a name, the system could recognize or profile an individual by these signatures (P2/P6). A particularly futuristic yet possible risk raised was the use of Artificial Intelligence (AI) on this data.

Augmented reality by design uses outward-facing cameras to map the real world. This means an AR device can record not only the user's surroundings but also other people in the vicinity without their knowledge. *"AR and Mixed Reality are most dangerous because they capture your environment... and can take images of other people without their consent if they're in your house,"* (P12). On the VR side, even though it's a simulated environment, the user's every action in that virtual space is observable by the platform. As one expert described, *"in VR... it's a totally new space. There is a system that sees you, tracks you in the* *virtual environment,"* like someone watching through a window at all times (P3). This can feel deeply invasive because the user's avatar and interactions (where they look, what they do, who they talk to) are being logged comprehensively.

Several respondents marked identity theft and avatar impersonation as a critical threat. Because avatars in the metaverse may look, move, or sound like the user stealing that data could let someone impersonate you convincingly. *"Identity theft is super easy if privacy fails... The moment I am streaming a large amount of data that allows others to recognize me as myself... the first thing I think of is impersonation,"* (P11).

Developers fear the metaverse could magnify the already intrusive targeted advertising of today's web. *"Let's be honest, there is no privacy now [on the web]. There's not gonna be privacy if this goes live in 3D... It'll be the same thing happening now, even more aggressive since everything is 3D and more imposing,"* (P10). They imagine marketers tracking *everything* a user does in VR where they go, what items they interact with, who they meet and using that to bombard the user with tailored ads inside the immersive environment. This could feel far more pervasive than banner ads on a phone screen because in VR/AR the ads could be integrated into one's very field of view and surroundings.

A concrete scenario given: *"You open the camera to map the room around you, and there's a Coca-Cola can... they can use that afterwards as an ad campaign – showing you items and products you like,"* (P2). Even more alarming, the rich data from XR could be leveraged not just to sell products but to shape opinions and behaviour. One participant further explained that the advertising argument to political and social manipulation: if a platform can infer a user's tendencies from nuanced 3D interaction data (much more than just clicks), it could *"curate what they see – articles, content – to shape socio-political views,"* (P10). This is analogous to the Facebook/Cambridge Analytica scandal (Cadwalladr, 2018; Kitchgaessner, 2018), but potentially on steroids: imagine a metaverse platform subtly adjusting a whole immersive world of content to nudge a user's beliefs, using knowledge of their psychological profile gleaned from biometrics.

In the interviews, participants debated whether VR or AR poses greater privacy risks, yet all agreed that XR significantly expands the data surface for potential privacy invasion. One participant argued AR is more dangerous because *"anyone can profile you and have an exact clone of yourself if they spend a little bit of time with you... it's more representative than a traditional game avatar,"* (P4), emphasizing how AR keeps you partially *"your real self"* in the system by making profiling easier. The safe conclusion was that all forms of XR possess these risks.

Developers conveyed that these risks challenge traditional definitions of privacy. One interviewee summarized, *"Your privacy in the metaverse is quite different... There need to be more things done on privacy"* (P8). *"In applications it's [just] data; in metaverse it's also images, movements, maybe senses. The principle is the same, but what is private is broader."* (P3). This calls for expanded thinking: designers and regulators must treat personal spatial and biometric data with the same care as we do personal information today.

A key factor discussed is the integration of AI in the metaverse, which developers see as both inevitable and double edged. *"AI needs to be present in the metaverse... it will be a necessity,"* (P11), but in the same breath warned that *"AI integration brings even higher data privacy needs than XR, to be honest."* (P11). One participant described how an AI could collect all the information about a user and environment and *"make a summary or generate content... but at the same time the AI keeps learning more about the person and stores all this information"* (P12).

## Discussion and Synthesis

The findings underscore the necessity of co-creative actions and strategies to ensure that privacy considerations are incorporated throughout the development of metaverse platforms. These privacy considerations must be accounted for during the requirements gathering phase, the design of the system, and the final implementation by developers. Collectively, the findings provide significant implications for technology ethics, both generally and within the specific context of metaverse platform creation. This research contributes to both theoretical and practical aspects of design and regulation, highlighting key insights that have emerged from interviews with experts in the field. Experts raised concerns about the nascent state of the metaverse, where end users often unaware of the implications of data privacy, may prioritize immersive user experiences over safeguarding their personal data. By deriving four key dimensions from expert's accounts, the study proposes a grounded framework that advances our understanding of privacy and ethics in immersive technologies. Based on the study's findings, we have outlined key practices and focus areas for metaverse experts and regulators in Table 2.

| Dimension | Practices for Embedding Privacy in Developer Workflows | Key Focus Areas |
|---|---|---|
| User Awareness & Empowerment | - Developing clear, user-friendly privacy settings.<br>- Educating users about their rights and tools.<br>- Building transparent privacy policies.<br>- Encouraging users to take control via default settings and easy-to-understand consent flows. | - User trust<br>-Transparency<br>-Consent and control<br>- Privacy education |
| Privacy by Design in Development | -Integrating privacy features into the core design & development process.<br>- Designing data minimization techniques.<br>- Using encryption and security measures by default.<br>- Conducting privacy impact assessments regularly. | -Data security<br>-Minimization of personal data<br>-Transparency in data use |
| Regulatory Governance | - Staying updated on privacy regulations.<br>- Implementing compliance monitoring processes.<br>- Creating an internal privacy policy that surpasses legal requirements.<br>- Engaging with external legal experts to ensure full regulatory compliance. | - Legal compliance<br>-Industry standards<br>- Risk management<br>-Proactive adaptation to new regulations |
| Novel Metaverse Privacy Risks | - Identifying biometric data risks, including consent and storage policies.<br>- Development of controls for AI-driven personalization and monitoring.<br>- Anticipating and protecting against identity theft and manipulation risks.<br>- Utilizing privacy safeguards for end users in metaverse environments. | -Biometric data protection<br>-Surveillance and manipulation risks<br>-Ethics in immersive technologies |

**Table 2. Synthesis Depicting Key Dimensions, Practices for Embedding Privacy in Metaverse Development**

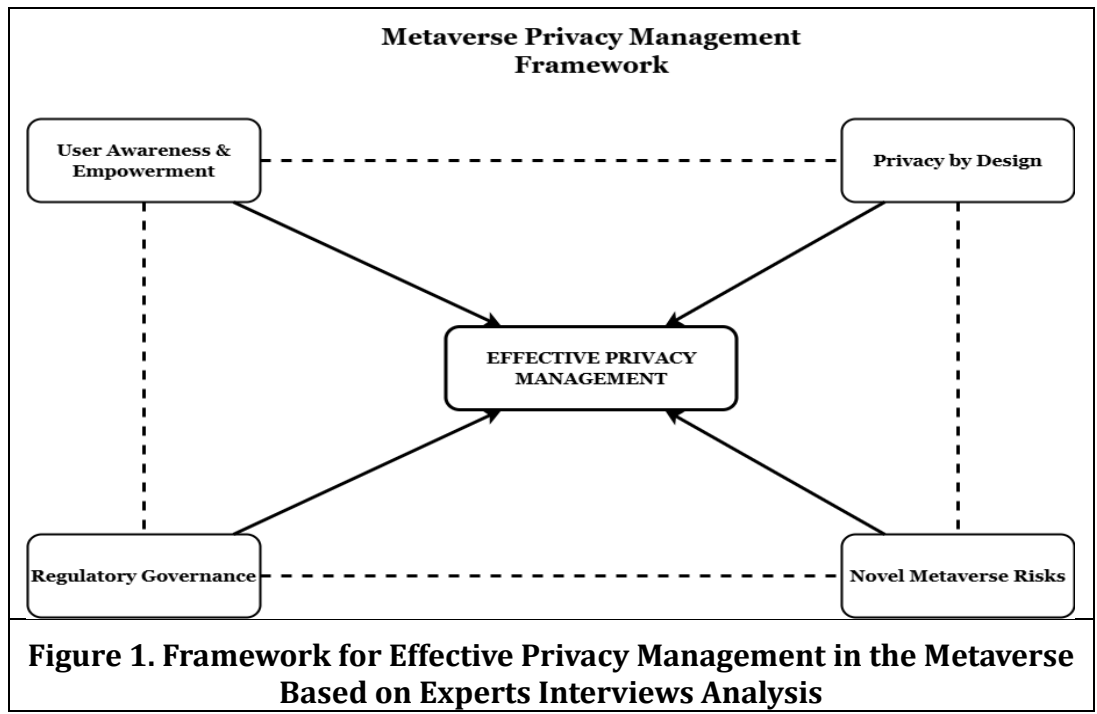

**Figure 1. Framework for Effective Privacy Management in the Metaverse Based on Experts Interviews Analysis**

Theoretically, this study expands existing privacy frameworks by integrating user behaviours, technical design, and regulation into a comprehensive model tailored to metaverse technologies. The proposed framework, as presented in Figure 1, enhances socio-technical systems theory, demonstrating that privacy behaviour within metaverse environments is shaped by users reduced awareness of data collection practices, thus escalating the privacy paradox.

Additionally, this research emphasizes the significance of privacy within technology acceptance and trust models, positing that perceived privacy protection is essential for the adoption of metaverse platforms. The findings suggest that traditional privacy theory is insufficient to address new types of personal data in the metaverse, such as behavioral biometrics and environmental data. These emerging data forms, which include users' movements, emotional responses, and interactions in virtual environments, present unique privacy challenges not covered by conventional frameworks. Therefore, in order to protect users in the metaverse effectively, privacy theory should be extended.

### Theoretical and Practical Implications

This research focuses on a critical issue regarding the responsible development of emerging technologies, particularly metaverse platforms. Since existing studies (Peukert et al., 2022; Wang et al., 2022) have focused on the commercial and user experience aspects of metaverse technologies. This paper emphasizes the importance of integrating ethical considerations into the development and deployment of these immersive technologies. By proposing a comprehensive framework (as shown in Figure 1), we highlight the connection of multiple ethical dimensions such as user awareness, empowerment, trust, privacy by design, developer responsibility, regulatory environment, and novel ethical risks associated with metaverse data.

Our framework advances existing literature by going beyond a siloed approach, where ethical concerns have frequently been handled as isolated concerns (Tiwari et al, 2025; Marabelli & Newell, 2023). For example, Tiwari et al. identified privacy concerns and ethical challenges as key issues in the metaverse, based on their systematic review. In contrast, our framework draws insights directly from the creators of metaverse technologies. We offer a holistic view of how ethical issues should be approached in the development of metaverse technologies, which contrasts with prior studies that have either focused on conceptual contributions or systematic reviews without providing empirical insights on the ethical implications (Nair et al., 2022a).

Furthermore, our research contributes to the growing body of literature on data-driven technologies by addressing concerns about privacy, security, and illegal access to sensitive user information (Dincelli &

Yayla, 2022; Bao et al., 2022). Although existing studies have raised concerns on the potential abuse of personal data (Di Pietro & Cresci, 2021), few of those studies have proposed specific frameworks to address these issues in immersive environments. Our work presents novel insights by identifying how the ethical risks related to data collection, sharing, and processing in the metaverse can be mitigated, thus bridging a gap in the current literature.

Future research can utilize this work as a baseline to comprehend and provide a more profound understanding of the complexities involved in metaverse creation/development. Researchers should not only explore individual ethical issues but also investigate the interplay between user rights, developer responsibilities, and regulatory frameworks. Apart from providing first-hand insights by the experts, our work also unfolds the findings of previous studies, such as Leenes (2008), Chen et al. (2022), Wang et al. (2022), and Al-Kfairy et al. (2025), and opens opportunities for future research on the ethical deployment of data-driven technologies in virtual environments. Practically, these findings offer actionable recommendations for stakeholders. For example, it is the responsibility of the entire software development team participating in the creation/development process of a metaverse platform to adopt privacy-by-design principles at every stage of metaverse development. They must ensure that privacy considerations are integrated as their standard software development practice rather than an optional feature.

Transparency in user experience is also critical, compelling companies to move beyond lengthy policies and instead embrace interactive, user-friendly privacy settings and educational initiatives. Regular audits of data flows, including third party components, are necessary to maintain privacy throughout the system. Organizations are encouraged to view privacy regulations as an opportunity for innovation, incorporating robust standards early in the development process to avoid expensive maintenance later.

Following the same lines, for regulators and policy makers, our research suggests the need to revise privacy definitions in order to account for new types of personal data, such as biometric and environmental data, while also upholding transparency measures tailored to metaverse technologies. on more practical level, there is a need to develop new privacy certification systems, and the encouragement of privacy-enhancing technologies (PETs) will be vital for establishing a trusted metaverse ecosystem. Moreover, empowering users through digital literacy campaigns will help create a more informed user base, driving demand for better privacy practices.

Companies can boost users trust, improve the user experience, and hence lay the groundwork for an ethical and sustainable metaverse by effectively addressing such sorts of interlinked challenges early in the design and implementation stages. These understandings offer a blueprint for ethical action, ensuring that privacy is integrated as a core element in both the design and governance of immersive technologies.

From our analysis of the interviews, we discovered several key practices for embedding privacy into developer workflows (as shown in Table 2). Current studies also advocate the adoption of these practices. For example, Allen et al (2021) highlight the value of including privacy from the scratch of the software development lifecycle (SDLC) via proactive privacy assessments and the integration of privacy and security measures. Similarly, Pallas et al. (2024) outline how tools like TILT can help in creating transparent privacy policies and make informed consent processes more accessible to users while also emphasizing the importance of encryption and security measures by default. Also, Haddar et al (2018) emphasized basic principles such as data minimization and the need for privacy compliance monitoring, providing practical techniques for enforcing privacy throughout the development process. Jointly, these existing studies stress a holistic method to privacy, from design to compliance, confirming that user privacy is prioritized and sustained around the SDLC.

### *Limitations and future research*

Our research also has some limitations that must be acknowledged. Firstly, it exclusively provides the perspectives of European metaverse experts on privacy in metaverse technologies. This research might serve as a baseline to explore the multifarious perspectives of metaverse developers from other regions, such as Asia and America, for the sake of better understanding how privacy concerns vary across different cultural and regulatory contexts. Additionally, while this paper focused primarily on developer's insights, further research could expand by interviewing other key stakeholders, such as end users, regulators and policymakers.

Gaining their perspectives would offer a more holistic understanding of privacy issues in the metaverse, especially regarding policy, regulation, and enforcement. Incorporating these directions with the previous suggestions, future studies could conduct quantitative research to assess how privacy features influence user trust and behaviour in metaverses across various regions and demographics. This would help to explore the privacy paradox in different cultural contexts and examine regional differences in privacy sensitivity. Longitudinal case studies could track metaverse development projects in various regions, documenting how privacy-by-design principles are implemented and identifying challenges developers face across different cultural or regulatory environments. Additionally, there is a need to evaluate the effectiveness of PETS, such as anonymization techniques and decentralized data processing methods, in metaverses. Comparative studies between regions could shed light on how privacy technologies are adopted and adapted in different markets.

Legal and policy analysis could focus on how existing privacy laws in regions such as the EU, US, and Asia apply to metaverse technologies, and propose new frameworks for privacy that consider regional regulatory perspectives. Building on this study, future research could also explore the feasibility of creating international standards for privacy protection in the metaverse. In terms of user centred design, future studies could investigate UX strategies tailored to the privacy concerns of diverse metaverses users, considering regional differences in consent mechanisms and transparency in data collection. Similarly, the development of global ethical frameworks for the metaverse could be informed by input from developers, regulators, users, and marginalized communities across different regions addressing privacy, consent, and identity issues from a broad, international perspective.

## Conclusion

This research highlights the multifaceted nature of privacy challenges in the metaverse, requiring collaboration across technical, human, and regulatory domains. By exploring developers' perspectives, we identified key areas towards ensuring user privacy in immersive environments that go beyond compliance and technology, focusing on user-centric design and proactive risk management. Achieving trust and ethical integrity in metaverses demands comprehensive strategies involving developers, users, organizations, and regulators. Our work contributes to privacy research by offering a framework tailored to metaverses, providing practical guidance for developers, organizations, and policymakers. This paper sets the stage for further efforts in building privacy-preserving, ethically responsible metaverse ecosystems.

## Acknowledgements

This research was funded by the European Union's Horizon Europe research and innovation programme "Training Young Researchers on Shaping Metaverse for Business and Social Value (AGORA)" under the Marie Skłodowska-Curie grant agreement No 101119937.

## References

Al-Kfairy, M., Alrabaee, S., Alfandi, O., Mohamed, A. T., & Khaddaj, S. (2025). Navigating ethical dimensions in the metaverse: Challenges, frameworks, and solutions. IEEE Access.

Allen, J., Hood, R., & Howell, K. (2021). Integrating Privacy Practices into the Software Development Lifecycle Transcript. Part 1: Keep Privacy at the Forefront; Collect Only Essential Information.

Ball, M. (2022). The metaverse will reshape our lives. Let's make sure it's for the better. 2022-08-08)[2023-02-28]. https://time. com/6197849/metaverse-future-matthew-ball.

Bao, X., Yu, J., & Shou, M. (2024). Exploring metaverse: affordances and risks for potential users. Information Technology & People.

Boughzala, I., de Vreede, G. J., & Limayem, M. (2012). Team collaboration in virtual worlds: Introduction to the special issue. Journal of the Association for Information Systems, 13(10), 7.

Cadwalladr, C. (2018). Facebook suspends data firm hired by Vote Leave over alleged Cambridge Analytica ties. The Guardian.

Chen, C., Li, Y., Wu, Z., Mai, C., Liu, Y., Hu, Y., ... & Zheng, Z. (2024). Privacy computing meets metaverse: Necessity, taxonomy and challenges. Ad Hoc Networks, 158, 103457.

Dolata, M., & Schwabe, G. (2023). What is the Metaverse and who seeks to define it? Mapping the site of social construction. Journal of Information Technology, 38(3), 239-266.

Dwivedi, Y. K., Hughes, L., Baabdullah, A. M., Ribeiro-Navarrete, S., Giannakis, M., Al-Debei, M. M., Dennehy, D., Metri, B., …, & Wamba, S. F. (2022). Metaverse beyond the hype: Multidisciplinary perspectives on emerging challenges, opportunities, and agenda for research, practice and policy. International Journal of Information Management, 66 (102542).

Dionisio, J. D. N., Iii, W. G. B., & Gilbert, R. (2013). 3D virtual worlds and the metaverse: Current status and future possibilities. ACM computing surveys (CSUR), 45(3), 1-38.

Di Pietro, R., & Cresci, S. (2021). Metaverse: Security and privacy issues. Third IEEE International Conference on Trust, Privacy and Security in Intelligent Systems and Applications (TPS-ISA).

Dincelli, E., & Yayla, A. (2022). Immersive virtual reality in the age of the Metaverse: A hybrid-narrative review based on the technology affordance perspective. The journal of strategic information systems, 31(2), 101717.

Falchuk, B., Loeb, S., & Neff, R. (2018). The social metaverse: Battle for privacy. IEEE technology and society magazine, 37(2), 52-61.

Floridi, L. (2020). The fight for digital sovereignty: What it is, and why it matters, especially for the EU. *Philosophy & technology*, *33*, 369-378.

Gioia, D. A., Corley, K. G., & Hamilton, A. L. (2013). Seeking qualitative rigor in inductive research: Notes on the Gioia methodology. Organizational research methods, 16(1), 15-31.

Hadar, I., Hasson, T., Ayalon, O., Toch, E., Birnhack, M., Sherman, S., & Balissa, A. (2018). Privacy by designers: software developers' privacy mindset. *Empirical Software Engineering*, *23*, 259-289.

Kitchgaessner, S. (2017). Cambridge Analytica used data from Facebook and Politico to help Trump. The Guardian, 26.

Khalid, M. I. (2025, September). Rethinking Digital Marketing Ethics in the Metaverse: A Framework for Responsible Engagement. In *Conference on e-Business, e-Services and e-Society* (pp. 362-377). Cham: Springer Nature Switzerland.

Leenes, R. (2008). Privacy in the metaverse: Regulating a complex social construct in a virtual world. In S. Fischer-Hubner, P. Duquenoy, A. Zuccato and L. Martucci (Eds.), *Ipip international federation for information processing* (pp. 95-112), Boston: Springer

McKinsey (2022a). Value creation in the metaverse - the real business of the virtual world.

McKinsey (2022b). Marketing in the metaverse: An opportunity for innovation and experimentation. https://www.mckinsey.com/business-functions/growth-marketing-and-sales/our-insights/marketing-in-the-metaverse-an-opportunity-for-innovation-and-experimentation

Marabelli, M., & Newell, S. (2023). Responsibly strategizing with the metaverse: Business implications and DEI opportunities and challenges. The Journal of Strategic Information Systems, 32(2), 101774.

McGonigal, J. (2011). *Reality is broken: Why games make us better and how they can change the world*. Penguin.

Nair, V., Guo, W., Mattern, J., Wang, R., O'Brien, J. F., Rosenberg, L., & Song, D. (2023). Unique identification of 50,000+ virtual reality users from head & hand motion data. arXiv preprint 2302.08927

Nair, V., Garrido, G. M., & Song, D. (2022). Exploring the unprecedented privacy risks of the metaverse. arXiv preprint 2302.08927.

Nissenbaum, H. (2004). Privacy as contextual integrity. *Wash. L. Rev.*, *79*, 119.

Peukert, C., Weinhardt, C., Hinz, O., & van der Aalst, W. M. (2022). Metaverse: How to approach its challenges from a BISE perspective. Business & Information Systems Engineering, 64(4), 401-406.

Pallas, F., Koerner, K., Barberá, I., Hoepman, J. H., Jensen, M., Narla, N. R., … & Zimmermann, C. (2024). Privacy engineering from principles to practice: A roadmap. *IEEE Security & Privacy*, *22*(2), 86-92.

Smith, H. J., Dinev, T., & Xu, H. (2011). Information privacy research: an interdisciplinary review. MIS quarterly, 989-1015.

Tiwari, M., Zhou, Y., Childs, A., Chang, L. Y., & Ferrill, J. (2025). Metaverse Policing: A Systematic Literature Review of Challenges and Recommendations. *Computers in Human Behavior*, 108591.

Wang, Y., Su, Z., Zhang, N., Xing, R., Liu, D., Luan, T. H., & Shen, X. (2022). A survey on metaverse: Fundamentals, security, and privacy. IEEE communications surveys & tutorials, 25(1), 319-352.

Zuboff, S. (2019, January). Surveillance capitalism and the challenge of collective action. In New labor forum (Vol. 28, No. 1, pp. 10-29). Sage CA: Los Angeles, CA: sage Publications.